\documentclass[preprint,12pt]{elsarticle}

\usepackage{amssymb}
\usepackage{amsthm}
\usepackage{amsmath}
\usepackage{color}

\journal{Applied Numerical Mathematics}

\newcommand{\Link}{Y}
\newcommand{\link}{y}
\newcommand{\Moment}{\Psi}
\newcommand{\moment}{\psi}
\newcommand{\alink}{\omega}
\newcommand{\Alink}{\Omega}
\newcommand{\slink}{x}
\newcommand{\Slink}{X}
\newcommand{\ad}{\text{ad}}

\newcommand{\SU}{\text{SU}}
\newcommand{\su}{\mathfrak{su}}
\newcommand{\group}{G}
\newcommand{\algebra}{\mathfrak{g}}
\newcommand{\ie}{i.~e.}

\newcommand{\C}{\mathbb{C}}
\newcommand{\R}{\mathbb{R}}
\newcommand{\N}{\mathbb{N}}

\newdefinition{rmk}{Remark}

\newtheorem{definition}{Definition}[section]
\newtheorem{lemma}[definition]{Lemma}

\begin{document}

\begin{frontmatter}



\title{Symmetric Partitioned Runge-Kutta Methods for Differential Equations on Lie Groups \tnoteref{t1}}
\tnotetext[t1]{This work was supported by the
Deutsche Forschungsgemeinschaft through the Collaborative Research Centre SFB-TR 55 `` Hadron
physics from Lattice QCD''.
}

\author{M.~Wandelt, M.~G\"unther, F.~Knechtli and M.~Striebel}

\address{Bergische Universit\"at Wuppertal, Fachbereich
Mathematik und Naturwissenschaften, D-42119 Wuppertal, Germany}
 \ead{\{wandelt,guenther,striebel\}@math.uni-wuppertal.de, knechtli@physik.uni-wuppertal.de}
\begin{abstract}
In this paper, we develop a higher order symmetric partitioned Runge-Kutta method for a coupled system of differential equations on Lie groups.
We start with a discussion on partitioned Runge-Kutta methods on Lie groups of arbitrary order. As symmetry is not met for higher orders, we generalize the method to a symmetric partitioned Runge-Kutta (SPRK) scheme.
Furthermore, we derive a set of coefficients for convergence order 4. The SPRK integration method can be used, for example, in simulations of quantum field theories. 
Finally, we compare the new SPRK scheme numerically with the St\"ormer-Verlet scheme, one of the state-of-the-art schemes used in this subject. 
 
%

\end{abstract}


\end{frontmatter}


\section{Introduction}
In the simulation of gauge theories in lattice Quantum Chromodynamics (QCD), for example, one is interested in calculating expectation values of certain operators. That means, very high dimensional ($10^7$ or more) integrals have to be evaluated. As this can not be done analytically in general, numerical methods are applied to get approximations to these integrals, \ie, the expectation values.

The Hybrid Monte Carlo (HMC) method \cite{duane_kennedy_etall:hmc,Gottlieb:1987mq} is widely used here. 
Alike in Monte Carlo integration, in HMC the integration is realized by averaging over evaluations of the integrand at certain, suitably chosen (importance sampling), values. In HMC, these points -- or samples -- are drawn from a combination of so-called Molecular Dynamics (MD) and Metropolis steps. In the former, starting from a suitable sample, a candidate for a next sample is derived from solving some differential equations. In the latter, it is checked whether this candidate suits or not, \ie, if it follows a certain distribution or not.

In this paper, we take a close look at the numerical integration of the differential equations in the MD step, arising in QCD problems. Commonly the Leapfrog (Stoermer-Verlet) scheme, Omelyan methods \cite{Omelyan2003272,Takaishi:2005tz} or splitting methods with multiple timescales a la Sexton-Weingarten \cite{Sexton:1992nu} are used. We formulate time-reversible higher order integrators that are based on implicit partitioned Runge-Kutta schemes and show that they allow for larger step-sizes than the Leapfrog method.

The paper is organized as follows: The equations of motion within the MD step
of HMC, together with the requirements to preserve the Lie group structure and time reversibility in numerical integration schemes,  are introduced in section 2. 
Partitioned Runge-Kutta (PRK) schemes are discussed in Section 3, based on Magnus expansion and Munthe-Kaas approach. 
They define numerical methods which preserve the Lie group structure. 
As this class of methods allows only for order lower or equal two, if in addition time reversibility has to be met, we generalize PRK methods to symmetric PRK (SPRK) methods which allow for higher order and derive a set of coefficients for a method of order 4 with 3 stages. 
The numerical results obtained in Section 5 show the efficiency of SPRK methods compared to the Leapfrog scheme at lower tolerances. It turns out that the integration measure (area) is not preserved with SPRK methods, i.e. they are not symplectic. This means that the determinant of the Jacobian has to be included in the Metropolis step. A conclusion and outlook to open question and future work is finally given in Section 6.

\section{Equations of Motion}\label{section:eqm}

We will not go into details for the origin of the differential equations to be solved in the MD step. For a detailed discussion we refer to \cite{Rothe_1992} and \cite{Kennedy:2006ax}. The dynamical system to be solved can be thought of as equations of motion, derived from some Hamiltonian operator $H(\Link,\Moment)$. In lattice QCD especially, these equations of motion form coupled systems of matrix differential equations of the form
\begin{subequations}
\label{eq:hamilton}
\begin{align}
\dot{\link}_\nu &= \phantom{-}\frac{\partial H\left(\Link,\Moment\right)}{\partial \moment_\nu} 
	=  
\moment_\nu \cdot \link_\nu,\label{eq:hamilton_link}\\
\dot{\moment}_\nu &= - \frac{\partial H\left(\Link,\Moment\right)}{\partial \link_\nu} = g_\nu(\Link), \qquad \text{for $\nu=1,\dotsc, n$}.\label{eq:hamilton_momentum}
\end{align}
\end{subequations}

Thereby, $\Link$ is a vector of $n$ elements $\link_1,\dotsc, \link_n$, each being an element of a matrix Lie group $\group$; the vector $\Moment$ comprises $n$ elements $\moment_\nu$ $(\nu=1,\dotsc,n)$, each being an element of the Lie algebra $\algebra$ associated to the Lie group $\group$. 

The coupled system \eqref{eq:hamilton} becomes an initial value problem (IVP) by prescribing initial values: $\link_\nu(0):= \link_{\nu,0} \in \group$ and $\moment_\nu(0):=\moment_{\nu,0} \in \algebra$ for $\nu=1,\dotsc,n$.

This IVP is usually solved by a numerical integration method on a time-grid $\{t_0=0, t_1, t_2\dotsc, t_{\text{end}} \}$. Here we let $(\Link_l, \Moment_l) = \Phi_h(\Link_{l-1}, \Moment_{l-1})$ represent a one-step method that computes an approximation $(\Link_l, \Moment_l) \approx (\Link(t_l), \Moment(t_l))$ at a time-point $t_l$, from values $(\Link_{l-1}, \Moment_{l-1})$ of the Lie group and Lie algebra elements at the time-point $t_{l-1}$. The progress in time is given by the step-size $h = t_l - t_{l-1}$. The accuracy of the method $\Phi_h$ is measured by the deviation $e_l=\| (\Link_l, \Moment_l) - (\Link(t_l), \Moment(t_l)\|$ with some suitable norm $\|\cdot\|$. 
The method is said to be of local order $p$ if $e_l = \mathcal{O}(h^{p+1})$ and $(\Link_{l-1}, \Moment_{l-1}) = (\Link(t_{l-1}), \Moment(t_{l-1}))$.

For efficiency, the order $p$ of the method should be preferably high, as this allows for large step-sizes $h$ to satisfy prescribed error tolerances. In addition,  
numerical schemes applied to the Lie group problem~\eqref{eq:hamilton} have to be equipped with the following properties:
{\renewcommand{\labelenumi}{(\alph{enumi})}
\begin{itemize}
\item The Lie group structure has to be preserved. That means,
approximations to $\Link$ and $\Moment$ have to reside in $\group$ and $\algebra$, element by element, respectively;
\item The integration scheme has to be symmetric. This is a consequence
of the detailed balance condition of the Markov process defined by the
HMC method 
 (see \cite{duane_kennedy_etall:hmc});
\item Detailed balance requires area preservation, i.e. a symplectic
integrator. If the area is not preserved by the integration, the
determinant of the Jacobian has to be included in the Metropolis
accept-reject step.
\end{itemize}
In lattice gauge theories, the state-of-the-art integration methods applied 
to \eqref{eq:hamilton} are the Leapfrog scheme of order 2 as well as Omelyan \cite{Omelyan2003272,Takaishi:2005tz} or splitting methods for higher orders \cite{Sexton:1992nu}.
In this paper, we develop an alternative to these schemes on the basis of partitioned implicit Runge-Kutta methods.

\section{Runge-Kutta Methods for Lie Group Problems}\label{sec:rklie}
Applying a numerical integration method directly to the coupled system \eqref{eq:hamilton}, it can not be guaranteed that the approximations to $\link_\nu$ (for $\nu=1,\dotsc,n$) are elements of the matrix Lie group $\group$, which is closed under matrix multiplication but not under summation.
%
Hence, measures have to be taken to preserve the Lie group structure in the numerical approximation. 

\subsection{The Magnus expansion}
The theorem of Magnus \cite{magnus:exp_solution} allows to transform the Lie group differential equation \eqref{eq:hamilton_link} to
\begin{align}
\label{eq:magnus}
 \dot \alink_\nu = d \exp_{\alink_\nu}^{-1}(\moment_\nu), \quad\text{for $\nu=1,\dotsc,n$},
\end{align}
with $\alink_\nu(t)\in\algebra$ and $\alink_\nu(0):=0$, \ie, to a differential equation in the corresponding Lie algebra $\algebra$. The way back from the Lie algebra $\algebra$ to the Lie group $\group$ is given by the mapping $\link_\nu(t)=\exp \left(\alink
_\nu (t)\right) \link_{\nu,0}$ where $\exp(\cdot)$ is the matrix exponential.

The central point in the transformation \eqref{eq:magnus} is $d \exp_{\alink_\nu}^{-1}$, the derivative of the inverse of the matrix exponential. This is given by the series
\begin{align}\label{eq:dexpinv}
 d \exp_{\alink_\nu}^{-1} \left(\moment_\nu\right) =\sum_{k \ge 0} \frac{B_k}{k!} ad_{\alink_\nu}^k \left(\moment_\nu\right).
\end{align}
Here, $B_k$ is the $k$-th Bernoulli number and $\ad_{\alink_\nu}$ is the adjoint operator, defined by $\ad_{\alink_\nu}(\moment_\nu):=[\alink_\nu,\moment_\nu]=\alink_\nu\moment_\nu - \moment_\nu\alink_\nu$ and $\ad_{\alink_\nu}^k(\moment_\nu) = [\alink_\nu, \ad_{\alink_\nu}^{k-1}(\moment_\nu)]$ with the convention $\ad_{\alink_\nu}^0(\moment_\nu) = \moment_\nu$. For a detailed discussion we refer to \cite{hairer_lubich_wanner:gni}.

In total, we record that the problem IVP \eqref{eq:hamilton} is equivalent to 
\begin{subequations}
\label{eq:mtran}
\begin{align}
\dot{\alink}_\nu &= \sum_{k=0}^{\infty} \frac{B_k}{k!} \ad_{\alink_\nu}^k(\moment_\nu), \label{eq:mtran_alink}\\
\dot{\moment}_\nu &= g_\nu(\Link)
\quad\text{with}\;\;\Link = (\link_\nu)_{\nu=1,\dotsc,n} \;\;\text{where}\;\;\link_\nu=\exp(\alink_\nu)\link_{\nu,0},\label{eq:mtran_moment}
\end{align}
\end{subequations}
with $\link_\nu(0):=\link_{\nu,0} \in \group$, $\moment_\nu(0):=\moment_{\nu,0} \in \algebra$ and $\alink_\nu(0):=0\in\algebra$ for $\nu=1,\dotsc,n$.

This transformed problem can now be solved directly by a Runge-Kutta method without destroying the Lie group structure: as the Lie algebra $\algebra$ is a vector space \cite{Cornwell_1997} both analytic solution $\alink_\nu(t), \moment_\nu(t)$ as well as approximations attained by a numerical integration scheme are elements of the Lie algebra $\algebra$. Furthermore, as for any $a\in \algebra$ the matrix exponential $\exp(a)$ is in the associated matrix Lie group $\group$, also $\link_\nu$ is in $\group$.

\subsection{The Munthe-Kaas approach}\label{section:MK}
Clearly, in practical computations the series in \eqref{eq:mtran_alink}, which is the expansion of $d\exp_{\alink_\nu}^{-1}$ given in equation \eqref{eq:dexpinv}, can not be evaluated. Instead, one has to truncate the series after some $q+1$ terms, \ie, sum up for $k=0$ to $k=q$.

Munthe-Kaas explains in \cite{Munthe-Kaas99highorder} how the truncation index $q$ can be chosen properly.
According to the observations made therein, a numerical integration method of local order $p$ applied to the Lie algebra problem \eqref{eq:mtran}, demands to take into account at least the first $p-1$ addends in the series. That means, the truncation index $q$ has to satisfy $q\ge p-2$.

By truncating the series in \eqref{eq:mtran_alink} at $k=q=p-2$, for a fixed $p\in \N$, the coupled system \eqref{eq:mtran} is formally replaced by the truncated IVP model
\begin{subequations}
\label{eq:mtrantrunc}
\begin{align}
\dot{\widehat{\alink}}_\nu &= \sum_{k=0}^{p-2} \frac{B_k}{k!} \ad_{\widehat{\alink}_\nu}^k(\widehat{\moment}_\nu), \label{eq:mtrantrunc_alink}\\
\dot{\widehat{\moment}}_\nu &= g_\nu(\widehat{\Link})
\quad\text{with}\;\;\widehat{\Link} = (\widehat{\link}_\nu)_{\nu=1,\dotsc,n} \;\;\text{where}\;\;\widehat{\link}_\nu=\exp(\widehat{\alink}_\nu)\widehat{\link}_{\nu,0},\label{eq:mtrantrunc_moment}
\end{align}
\end{subequations}
with $\widehat{\link}_\nu(0):={\link}_{\nu,0} \in \group$, $\widehat{\moment}_\nu(0):={\moment}_{i,0} \in \algebra$ and $\widehat{\alink}_\nu(0):=0\in\algebra$ for $\nu=1,\dotsc,n$.

We can now apply a numerical integration scheme of local order $p$. This yields approximations
$(\link_{\nu,1}, \alink_{\nu,1}, \moment_{\nu,1})$ to the exact solution $(\widehat{\link}_\nu,\widehat{\alink}_\nu,\widehat{\moment}_\nu)(h)$ of the truncated model \eqref{eq:mtrantrunc} at time-point $t_1=0+h$ of local order $p$. That means, the approximations satisfy $\|\widehat{\moment}_{\nu}(h) - {\moment}_{\nu,1}\| = \mathcal{O}(h^{p+1})$, correspondingly for $\widehat{\link}_{\nu}$ and $\widehat{\alink}_\nu$, for $\nu=1,\dotsc,n$.

The central statement of Munthe-Kaas \cite{Munthe-Kaas99highorder} is that $(\link_{\nu,1}, \alink_{\nu,1}, \moment_{\nu,1})$ is also an approximation  of local order $p$ to the exact solution $({\link}_\nu,{\alink}_\nu,{\moment}_\nu)(h)$ of the original problem \eqref{eq:mtran}. Or, the other way around: a method to compute approximations of order $p$ to the exact solution of the Lie algebra problem \eqref{eq:mtran} consists in applying a numerical integration scheme of order $p$ to the truncated model \eqref{eq:mtrantrunc}. 

In the latter problem, \ie, in the dynamical system \eqref{eq:mtrantrunc}, we skip the $\widehat{\phantom{\alink}}$-notation and use in the following the condensed formulation
\begin{subequations}
\label{eq:ode-system}
\begin{align}
\dot \Alink &= f(\Alink,\Moment), \label{eq:ode-system.a} \\
\dot \Moment &=  g(\Link) \;\; \text{with} \;\; \Link= \exp(\Alink)\Link_0,
\label{eq:ode-system.b}
\end{align}
\end{subequations}
where
\begin{align*}
\Link(t) = (\link_\nu(t))_{\nu=1,\dotsc,n}\,,\;\Moment(t) = (\moment_\nu(t))_{\nu=1,\dotsc,n}\,,\;\text{and}\;\Alink(t) = (\alink_\nu(t))_{\nu=1,\dotsc,n}\,,
\end{align*}
and 
\begin{align*}
f(\Alink,\Moment) &:= \left(f_\nu(\alink_\nu,\moment_\nu)\right)_{\nu=1,\dotsc,n} \quad\text{with}\;\;f_\nu(\alink_\nu,\moment_\nu)= \sum_{k=0}^{p-2} \frac{B_k}{k!}\ad_{\alink_\nu}^k(\moment_\nu)\,,\\
g(Y)&:=\left(g_\nu(Y)\right)_{\nu=1,\dotsc,n}\;\;\text{and}\;\;
\exp(\Alink)\Link_0 := \left(\exp(\alink_\nu)\link_{\nu,0}\right)_{i=1,\dotsc,n}
\end{align*}
with initial values $\Link(0):=\Link_0:=(\link_{\nu,0})_{\nu=1,\dotsc,n}$, $\Moment(0):=\Moment_0:=(\moment_{\nu,0})_{\nu=1,\dotsc,n}$ and $\Alink_0:=(0)_{\nu=1,\dotsc,n}$.

\subsection{Partitioned Runge-Kutta methods}\label{sec:prk}
To solve the coupled dynamical system \eqref{eq:ode-system}, we apply a partitioned Runge-Kutta (PRK) method \cite{hairer_norsett_wanner:solving1} with $s$ stages and coefficients $b_i, \widehat{b}_i, \alpha_{ij}, \widehat{\alpha}_{ij}$ for $i,j=1, \dotsc, s$. Starting from $t=0$ we first compute approximations $\Alink_1$ and $\Moment_1$ at the time-point $t=h$ by
\begin{subequations}
\label{eq:prk_ode_lie}
\begin{align}
\Alink_1 &= \Alink_0 + h\sum_{i=1}^sb_i K_i, & \Moment_1 &= \Moment_0 + h\sum_{i=1}^s \widehat{b}_i L_i,\label{eq:prk_ode_lie_approx}\\
\intertext{with increments $K_i$ and $L_i$ for $i=1,\dotsc,s$ defined by}
K_i &= f\left(\bar{\Alink}_i, \bar{\Moment}_i\right), & L_i&=g\left(\bar{\Link}_i\right),
\label{eq:prk_ode_lie_increments}
\end{align}
where $\bar{\Alink}_i$, $\bar{\Moment}_i$ and $\bar{\Link}_i$ are internal stages given by
\begin{align}
\bar{\Alink}_i = \Alink_0 +h\sum_{j=1}^s \alpha_{ij}K_j, \quad
\bar{\Moment}_i = \Moment_0+h\sum_{j=1}^s \widehat{\alpha}_{ij}L_j,\quad 
\bar{\Link}_i = \exp\left(\bar{\Alink}_i\right)\Link_0.
\label{eq:prk_ode_lie_stages}
\end{align}
Finally, we get an approximation $(\link_{\nu,1}, \moment_{\nu,1})$ of local order $p$ to the exact solution $(\link_\nu(t), \moment_\nu(t))$, one time-step $h$ ahead, \ie, at $t=0+h$ by
\begin{align}
\link_{\nu,1} = \left(\Link_1\right)_{\nu}\;\;\text{with}\;\; \Link_1:=\exp\left(\Alink_1\right)\Link_0 \quad \text{and} \quad \moment_{\nu,1} = \left(\Moment_1\right)_{\nu}.
\end{align}
 \end{subequations}
 
The coefficients $b_i, \widehat{b}_i, \alpha_{ij}, \widehat{\alpha}_{ij}$ for $i,j=1, \dotsc, s$ steer the behavior of the method and have to be chosen properly. Conditions, the coefficients have to satisfy to attain an approximation of local order $p$, are found by series expansions of the approximations produced by the method \eqref{eq:prk_ode_lie} and the exact solution of the truncated problem \eqref{eq:mtrantrunc} in powers of $h$, followed by a comparison of the series' coefficients. For an accuracy of order $2$, for example, the order conditions for $p=1$ and $p=2$ as stated in Tab.~\ref{table:order2} have to be fulfilled. 

\begin{table}[htb]
{\renewcommand{\arraystretch}{1.5}
\begin{center}
\begin{tabular}{|l|c|c|}
  \hline
  p & $\Alink(t)$ & $\Moment(t)$\\[1ex]
  \hline
  $1$ & $\sum_i b_i = 1$
      & $\sum_i \widehat{b}_i =1$\\[2ex]
  $2$ & $\sum_{i} b_i \widehat{\alpha}_{i} = 1/2$
      & $\sum_{i} \widehat{b}_i \alpha_{i} = 1/2$\\[1ex]
  \hline 
\end{tabular}
\end{center}}
\caption{Order conditions for the PRK method \eqref{eq:prk_ode_lie}, see \cite{hairer_lubich_wanner:gni} or  \cite{hairer_norsett_wanner:solving1}. \label{table:order2}}
\end{table}

\section{Symmetric PRK methods for Lie group problems}\label{section:sprk}
{Symmetry}, frequently termed {\em time-reversibility}, is closely related to the {\em adjoint} of a method. The {\em adjoint method} $\Phi_h^\star$ of a method $\Phi_h$ is the inverse of the original method with reversed time-step $-h$, \ie 
\begin{equation}
\label{eq:adjoint}
\Phi_h^\star := \Phi_{-h}^{-1}.
\end{equation}
A numerical integration scheme $\Phi_h$ is symmetric if it equal to its adjoint $\Phi_h^\star$, \ie, $\Phi_h = \Phi_h^\star$. The specifications for the adjoint method are found by exchanging the roles of the initial value and the approximation, reversing the time-step and dissolving this system again for the approximation. For a detailed discussion of the procedure, we refer to  \cite{hairer_lubich_wanner:gni}. 

\subsection{Symmetric Lie group PRK method}\label{sec:sym_prk}
When deriving the adjoint method of the PRK method \eqref{eq:prk_ode_lie}, however, we have to pay attention to the Lie group structure. Therefore, we state some more details.

As described, we exchange $(h$,$\Link_0$,$ \Moment_0$,$ \Link_1$,$ \Moment_1)$ in \eqref{eq:prk_ode_lie} by $(-h$,$\Link_1$,$ \Moment_1$,$ \Link_0$,$ \Moment_0)$  along with replacing $\Alink_1$ with $-\Alink_1$ and then resolve for $(\Link_1, \Moment_1)$.

The arising instructions can be stated as a PRK method, similar to the original method \eqref{eq:prk_ode_lie}. The coefficients $b_i^\star$, $\widehat{b}_i^\star$, $\alpha_{ij}^\star$, $\widehat{\alpha}_{ij}^\star$ for $i,j=1,\dotsc,s$ are connected to the coefficients of the original method via 
\begin{align*}
\widehat{b_i}^\star :=\widehat{b}_{s+1-i},\;\text{and}\;\;
\widehat{\alpha}_{ij}^\star := \widehat{b}_{s+1-j} - \widehat{\alpha}_{s+1-i,s+1-j},
\end{align*}
and for ${b}_i$ and ${\alpha}_{ij}$ in equal measure.
There are, however, two important differences. The increments $K_i$ in the adjoint method are given as
\begin{subequations}\label{eq:prk_adjoint}
\begin{align}
\label{eq:prk_adjoint_increment}
K_i = f(\widetilde{\Alink}_i, \bar{\Moment}_i^\star), \;\text{with}\;\;\widetilde{\Alink}_i:=h\sum_{j=1}^s (-\alpha_{s+1-i,s+1-j}) K_j.
\end{align}
$\bar{\Moment}_i^\star$ is defined in the obvious way.
Furthermore, the internal stages for the links are defined by
\begin{align}
\label{eq:prk_adjoint_stage}
\bar{\Link}_i^\star = \exp(\widetilde{\Alink}_i) \cdot \exp(\Alink_1) \Link_0.
\end{align}
\end{subequations}

The numerical scheme \eqref{eq:prk_ode_lie} is symmetric if it coincides with its adjoint method. That means, the approximations $(\Link_1, \Moment_1, \Alink_1)$ produced  and, therefore, the increments $(K_i, L_i)$ within $\Phi_h$ and $\Phi_h^\star$ have to agree. 

Immediately, the conditions 
\begin{subequations}\label{eq:prk_symmetry_conditions}
\begin{align}
b_i & = b_{s+1-i},\label{eq:prk_symmetry_conditions_b}\\
\widehat{b}_i &= \widehat{b}_{s+1-i},\label{eq:prk_symmetry_conditions_bhat}\\
\widehat{\alpha}_{ij} &= \widehat{b}_{s+1-j} -\widehat{\alpha}_{s+1-i,s+1-j}\label{eq:prk_symmetry_conditions_alphahat}
\intertext{for all $i,j = 1,\dotsc,n$ become clear. Considering the increments $L_i$, we see that $\bar{\Moment}_i = \bar{\Moment}_i^\star$ needs to hold. However, we can only derive a condition for the coefficients if $\widetilde{\Alink}_i$ and $\Alink_1$ in \eqref{eq:prk_adjoint_stage} commutate. The condition imposed then reads} 
{\alpha}_{ij} &= {b}_{s+1-j} -{\alpha}_{s+1-i,s+1-j}.\label{eq:prk_symmetry_conditions_alpha}
\intertext{$\widetilde{\Alink}_i$ and $\Alink_1$ commutate if $b_{s+1-j}$ is some multiple of $-\alpha_{s+1-i,s+1-j}$, \ie, if} 
b_j &= d_i \cdot \alpha_{ij} \;\text{with some $d_i\in\R$.}\label{eq:prk_symmetry_conditions_commutate}
\intertext{Finally, we consider the increments $K_i$. Comparing  \eqref{eq:prk_adjoint_increment} and \eqref{eq:prk_ode_lie_increments} yields the condition}
\alpha_{ij} &= -\alpha_{s+1-i,s+1-j}.\label{eq:prk_symmetry_conditions_cut}
\end{align} 
\end{subequations}
The conditions (\ref{eq:prk_symmetry_conditions_b}-d) are the usual symmetry conditions for PRK methods (see \cite{hairer_lubich_wanner:gni}). The conditions (\ref{eq:prk_symmetry_conditions_commutate},f) arise from the Lie group nature of the problem. 

The symmetry conditions \eqref{eq:prk_symmetry_conditions_alpha} and \eqref{eq:prk_symmetry_conditions_cut} imply $b_i=0$ for all $i=1,\dotsc,n$. Clearly, this is a contradiction to the basic order-1 condition $\sum_i b_i =1$. 

However, this contradiction does not appear if the local order $p$ of the method is $2$ at most. In this case, \eqref{eq:mtrantrunc_alink} simplifies to $\dot{\alink}_\nu=\moment_\nu$. Hence, the function $f$ only depends on $\Moment$, \ie, $f=f(\Moment)$. Consequently, the symmetry condition \eqref{eq:prk_symmetry_conditions_cut} vanishes, such that the conflict between order and symmetry condition disappears.

This statement is summarized in the following
\begin{lemma}\label{lemma:symmetry}
The PRK scheme \eqref{eq:prk_ode_lie} can only be symmetric if $p \le 2$.
\end{lemma}

Indeed, there is a symmetric partitioned Runge-Kutta scheme for Lie group problems.
The {\em St\"ormer-Verlet scheme}, also known as {\em Leapfrog method} in various applications, can be interpreted as partitioned Runge-Kutta scheme of type \eqref{eq:prk_ode_lie} with coefficients given by the
extended Butcher tableaus (see \cite{hairer_lubich_wanner:gni}) 
{\renewcommand{\arraystretch}{1.4}
\begin{align*}
\begin{array}{c|c}
\alpha&A\\\hline 
&b
\end{array} \;\;=\;\; 
\begin{array}{c|cc}
0&0&0\\
1&0.5&0.5\\\hline
&0.5&0.5
\end{array}\quad\text{and}\quad
\begin{array}{c|c}
\widehat{\alpha}&\widehat{A}\\\hline 
&\widehat{b}
\end{array} \;\;=\;\; 
\begin{array}{c|cc}
0.5&0.5&0\\
0.5&0.5&0\\\hline
&0.5&0.5,
\end{array}
\end{align*}}

where $A=(\alpha_{ij})_{i,j=1,\dotsc,s}$, $b=(b_1,\dotsc,b_s)$ and $\alpha=(\alpha_1,\dotsc,\alpha_s)^T$ with $\alpha_i=\sum_j \alpha_{ij}$, and in the same way for $\widehat{A},\widehat{b},\widehat{\alpha}$.

It is easily checked that these coefficients define a method of order two applied to the original system and
satisfy the conditions \eqref{eq:prk_symmetry_conditions} with $d_1 = 0$ and $d_2=0.5$ in equation \eqref{eq:prk_symmetry_conditions_commutate}. Therefore, the St\"ormer-Verlet scheme is a symmetric integrator.

\subsection{Higher Order Symmetric Partitioned RK Methods}
To allow for higher order symmetric methods, we need more flexibility in the stage vector
$\bar{\Link}_i$. We get this by introducing additional coefficients $\gamma_{i,j}$ for $i,j=1,\dotsc,s$ and replacing the links' internal stages $\bar{\Link}_i$ in \eqref{eq:prk_ode_lie_stages} with
\begin{equation}
\label{eq:replace_link_versionA}
\begin{split}
 \bar{\Link}_i &= \exp\left(\bar{\Slink}_i\right)\Link_0 = \left(\exp(\bar{\slink}_{\nu,i})  \cdot \link_{\nu,0} \right)_{\nu=1,\ldots,n} ,\\
\text{where}\;\;  \bar{\Slink}_{i} &= \left(\bar{\slink}_{\nu,i}\right)_{\nu=1,\ldots,n} = h\sum_{j=1}^s \gamma_{i,j} K_j. 
\end{split}
\end{equation}

By this, the aforementioned contradiction in the conditions vanishes because \eqref{eq:prk_symmetry_conditions_cut} and \eqref{eq:prk_symmetry_conditions_commutate}
are replaced in a first step by
\begin{subequations}\label{eq:PRK_symmetry_conditions_new}
\begin{align}
  \gamma_{i,j} &= b_{s+1-j} - \gamma_{s+1-i, s+1-j} \label{eq:SPRK_syym_cond_slink}\\
  \text{and}\quad
  b_j &= d_i \cdot \gamma_{i,j}. \label{eq:SPRK_syym_cond_slink_comm}
\end{align}
\end{subequations}

However, deriving the order conditions for this adapted PRK scheme, it turns out that the condition \eqref{eq:SPRK_syym_cond_slink_comm} for symmetry leads to a conflict with conditions for convergence orders $p>2$. 
The problem is caused by the term $\exp(\Alink_1)$ that enters the definition of the internal stage variable $\bar{\Link}_1^\star$ of the adjoint method. The coefficients $b_j$ have to be the multiple of some other method's coefficients to guarantee the commutativity of $\Alink_1$ with $\widetilde{\Alink}_i$ in \eqref{eq:prk_adjoint_stage} or with $\widetilde{\Slink}_i = h\sum_j (-\gamma_{s+1-i,s+1-j}) K_j$ if the variant \eqref{eq:replace_link_versionA} is used.

As a solution, this necessity -- and therefore the condition \eqref{eq:prk_symmetry_conditions_commutate} and \eqref{eq:SPRK_syym_cond_slink_comm}, respectively -- can be avoided by a further reformulation of the stage vector $\bar{\Link}_i$. Finally, we replace the link's internal stages in \eqref{eq:prk_ode_lie_stages} with

\begin{equation*}
\begin{split}
 \bar{\Link}_i &= \exp\left(\bar{\Slink}_i\right)\exp\left(\tfrac{1}{2}\Alink_1\right)\Link_0 = \left(\exp(\bar{\slink}_{\nu,i})  \cdot \exp(\bar{\alink}_{\nu,1})\cdot  \link_{\nu,0} \right)_{\nu=1,\ldots,n} ,\\
\text{where}\;\;  \bar{\Slink}_{i} &= \left(\bar{\slink}_{\nu,i}\right)_{\nu=1,\ldots,n} = h\sum_{j=1}^s \gamma_{i,j} K_j. 
\end{split}
\end{equation*}

and arrive at the symmetric partitioned Runge-Kutta scheme (SPRK)

\begin{subequations}
\label{eq:sprk_ode_lie}
\begin{align}
\Alink_1 &=  h\sum_{i=1}^sb_i K_i, & \Moment_1 &= \Moment_0 + h\sum_{i=1}^s \widehat{b}_i L_i, & \Link_1=\exp(\Alink_1)\Link_0
\intertext{with increments $K_i$ and $L_i$ for $i=1,\dotsc,s$ defined by}
K_i &= f\left(\bar{\Alink}_i, \bar{\Moment}_i\right), & L_i&=g\left(\bar{\Link}_i\right),
\label{eq:sprk_ode_lie_increments}
\end{align}
{and internal stages defined as}
\begin{align}
\bar{\Alink}_i &= h\sum_{j=1}^s \alpha_{ij}K_j, &
\bar{\Moment}_i &= \Moment_0+h\sum_{j=1}^s \widehat{\alpha}_{ij}L_j,\\
\bar{\Link}_i &=\exp\left(\bar{\Slink}_i\right)\exp\left(\tfrac{1}{2}\Alink_1\right)\Link_0, &\bar{\Slink}_{i} &= h\sum_{j=1}^s \gamma_{i,j} K_j,
\label{eq:sprk_ode_lie_stages}
\end{align}
 \end{subequations}

with coefficients $b_i, \widehat{b}_i, \alpha_{ij}, \widehat{\alpha}_{ij}, \gamma_{ij}$ for $i,j=1, \ldots, s$.

The SPRK method \eqref{eq:sprk_ode_lie} is not a "standard" partitioned RK method. Hence, we have to compute both symmetry and order conditions for this method.

\subsubsection{Symmetry conditions}

We first determine the conditions for symmetry that the method's coefficients have to fulfill. For this, we determine the adjoint method of the method \eqref{eq:sprk_ode_lie} and carry out the same steps we described in Sec.~\ref{sec:sym_prk} for the method at hand.

Again, the adjoint to the SPRK method can be formulated as a partitioned RK method with coefficients $b_i, \widehat{b}_i, \alpha_{ij}, \widehat{\alpha}_{ij}$ (like in the adjoint for the PRK method \eqref{eq:prk_ode_lie}) plus coefficients $\gamma_{ij}^\star:= -\gamma_{s+1-i,s+1-j}$ for $i,j=1,\dotsc,s$.

We recall that the root for the problems we recorded was the computation of the PRK adjoint method's increments $K_i$ and the internal stages $\bar{\Link}_i^\star$ as stated in \eqref{eq:prk_adjoint_increment} and \eqref{eq:prk_adjoint_stage}, respectively. 
Within the adjoint method of the SPRK scheme, the $K_i$'s are computed in the same way \eqref{eq:prk_adjoint_increment} but the internal stages $\bar{\Link}_i^\star$ now amount to 
\begin{align}\label{eq:sprk_adjoint}
\begin{aligned}
\bar{\Link}_i^\star &= \exp\left(\bar{\Slink}_i^\star\right)\cdot\exp\left(-\tfrac{1}{2}\Alink_1\right) \cdot \Link_1 \\
&= \exp\left(\bar{\Slink}_i^\star\right)\cdot\exp\left(\tfrac{1}{2}\Alink_1\right) \cdot \Link_0,
\end{aligned}
\end{align}
with $\bar{\Slink}_i^\star$ defined in the obvious way (as $h\sum_j \gamma^\star_{ij} K_j$).

Finally, the symmetry conditions, which the method's coefficients have to satisfy, are 
\begin{align}
\label{eq:sprk_symmetry_conditions}
\begin{aligned}
b_i &= b_{s+1-i}\;\; \text{and}\;\; \widehat{b}_i = \widehat{b}_{s+1-i},\\
\alpha_{ij} &= -\alpha_{s+1-i,s+i-j}, \\
\widehat{\alpha}_{ij} &= \widehat{b}_{s+1-j} - \widehat{\alpha}_{s+1-i,s+1-j},\\
\gamma_{ij}&= -\gamma_{s+1-i,s+1-j}
\end{aligned}
\end{align}
for all $i,j=1,\dotsc,s$.

Note that the absence of any condition like \eqref{eq:prk_symmetry_conditions_commutate} or \eqref{eq:SPRK_syym_cond_slink_comm} is due to similar form of the specifications of $\bar{\Link}_i$ and $\bar{\Link}_i^\star$ in \eqref{eq:sprk_ode_lie_stages} and \eqref{eq:sprk_adjoint}, respectively. Because of that, $\frac{1}{2}\Alink_1$ does not have to commutate, neither with $\bar{\Slink}_i$ nor with $\bar{\Slink}_i^\star$. 

Furthermore, the symmetry conditions do not conflict with the order conditions that we derive now.

\subsubsection{Order conditions}

As explained by Munthe-Kaas, the SPRK method \eqref{eq:sprk_ode_lie} is of local order $p$ with respect to the Lie group differential equation \eqref{eq:hamilton} (see Sec.~\ref{sec:rklie}) if 
\begin{equation}\label{eq:convergence_order_sprk}
\|\Link(t_0+h) -\Link_1\| = \mathcal{O}(h^{p+1}) \quad \text{and}\quad \|\Moment(t_0+h) -\Moment_1\| = \mathcal{O}(h^{p+1}),
\end{equation}
where $\Link_1$ and $\Moment_1$ are the approximations to exact solution $\Link(t_0+h), \Moment(t_0+h)$ of the suitably truncated problem \eqref{eq:ode-system} after one step.

As within the SPRK method, the approximation to the links $\Link_1$ arises from evaluating the matrix exponential (we assume here that we can evaluate this exactly), which is Lipschitz on every closed interval, it suffices to demand 
\begin{equation*}
\|\Alink(t_0+h) -\Alink_1\| = \mathcal{O}(h^{p+1}) \quad \text{and}\quad \|\Moment(t_0+h) -\Moment_1\| = \mathcal{O}(h^{p+1}).
\end{equation*}
As mentioned in Sec.~\ref{sec:prk}, expanding $(\Alink_1,\Moment_1)$ and $(\Alink(t_0+h),\Moment(t_0+h))$ to power series in $h$ leads to order conditions for the method's coefficients by matching the series' coefficients of the exact solution and its corresponding approximation. In constructing RK schemes, the use of some B-series' approach \cite{hairer_wanner:solving2,hairer_norsett_wanner:solving1} is common. However, to our knowledge, there is not yet a theory available that fits to the situation reported in this paper. Therefore, for the time being, we apply standard Taylor series expansion. 

We are concerned with a non-abelian, \ie, non-commutative, matrix Lie algebra $\algebra$.
Therefore, the Taylor expansion and the nature of the order condition deviate from the standard, \ie, abelian case. We recognize several issues:
\begin{itemize}
\item The right hand side \eqref{eq:ode-system.a}, \ie, the function $f$, depends on the maximal order $p$ we aim to attain. Each increase in order adds an additional $k$-th order commutator $\text{ad}_{\alink_\nu}^k(\moment_\nu)$. As a consequence, additional (w.~r.~t. the abelian case) order conditions appear.
\item The definition of the internal stages $\bar{\Link}_i$ in \eqref{eq:sprk_ode_lie_stages}, where the approximation $\Alink_i$ is included, is a non-standard RK formulation. This and the non-commutativity of $\bar{\Slink}_i$ and $\Alink_1$ also leads to further order conditions.
\end{itemize}

As the actual computation of the Taylor series's expansion in this case is tedious, we skip details. Instead, we just state the order conditions up to order $p=3$ in Tab.~\ref{table:order3}. Note that, due the maximum order $p=3$, the function $f$ had to be chosen as:
\begin{equation*}
f(\Alink,\Moment) = \Alink - \frac{1}{2}\left[ \Alink,\Moment\right].
\end{equation*}

\begin{table}[htb]
\begin{center}\renewcommand{\arraystretch}{1.5}
\begin{tabular}{|l|c|c|}
  \hline
  $p$& $\Alink(t)$ & $\Moment(t)$\\[1ex]
  \hline
  $1$ & $\sum_i b_i = 1$
      & $\sum_i \widehat{b}_i =1$\\[2ex]
  $2$ & $\sum_{i} b_i \widehat{\alpha}_{i} = 1/2$
      & $\sum_{i,j} \widehat{b}_i \left( \gamma_{i} +  b_j/2 \right) = 1/2$\\[2ex]
  $3$ & $\sum_{i,j} b_i \left(\alpha_{i} \widehat{\alpha}_{i} - \alpha_{ij} \widehat{\alpha}_{j} \right)= 1/6$
      & $\sum_{i,j} \widehat{b}_i \left(\gamma_{i} + b_j/2 \right)^2 = 1/3$\\[2ex]
      & $\sum_{i,j} b_i \widehat{\alpha}_{ij} \left(\gamma_{j} +  \sum_k  b_k/2 \right) = 1/6$
      & $\sum_{i,j} \widehat{b}_i \left(\gamma_{ij} + b_j/2  \right) \widehat{\alpha}_{j} = 1/6$\\[1ex]
  \hline 
  \multicolumn{3}{l}{with $\gamma_i: = \sum_j \gamma_{ij}$ and $\alpha_i: = \sum_j \alpha_{ij}$}
\end{tabular}
\end{center}
\caption{Order conditions for the symmetric partitioned Runge-Kutta (SPRK) method which is stated in \eqref{eq:sprk_ode_lie}.\label{table:order3}}
\end{table}

The following set of coefficients solves both the order conditions given in Tab.~\ref{table:order3} and the symmetry conditions specified in \eqref{eq:sprk_symmetry_conditions} and, hence, defines a SPRK method of order $p=3$ with $s=3$ stages:
\begin{align}\label{eq:sprk_coeffs}
\begin{aligned}
\alpha_{21} &= \gamma_{11} =-\sqrt{3}/6,&
\alpha_{23}  &= \gamma_{33} = \sqrt{3}/6, &
b_2 &= 1,\\
\widehat{\alpha}_{11} & = (3+\sqrt{3})/6,&
\widehat{\alpha}_{21} &= (3+\sqrt{3})/12, &
\widehat{b}_1& = \widehat{b}_3 = 1/2,\\
\widehat{\alpha}_{23} &= (3-\sqrt{3})/12,&
\widehat{\alpha}_{31} &= 1/2,&\\
\widehat{\alpha}_{33} &= -\sqrt{3}/{6}.
\end{aligned}
\end{align}
Here, coefficients that are not mentioned explicitly are $0$.
It is known in the literature \cite{hairer_lubich_wanner:gni} that symmetric Runge-Kutta schemes of order higher than 2 as well as partitioned ones for problems of the type \eqref{eq:ode-system} can not be explicit. 
Indeed, the coefficients \eqref{eq:sprk_coeffs} define an implicit SPRK method. 
It follows that the equations (\ref{eq:sprk_ode_lie}b-d) for the increments $K_i$ and $L_i$ of the SPRK scheme are implicit. We solve them through a fixed-point iteration.

\subsection{Global error}
Recall that the SPRK method \eqref{eq:sprk_ode_lie} is a numerical scheme that is constructed such that it produces approximations of local order $p$ with respect to the exact solution of the truncated IVP \eqref{eq:ode-system}. But, due to the argumentation of Munthe-Kaas (see Sec.~\ref{section:MK}), the results are also approximation of local order $p$ w.~r.~t. the original problem, \ie, the equations of motion \eqref{eq:hamilton}. That means, \eqref{eq:convergence_order_sprk} holds also for the exact solution $(\Link(t_0+h),\Moment(t_0+h))$ of \eqref{eq:hamilton}.

We now investigate the global error, \ie, the deviation between the exact solution $(\Link(t_0+N\cdot h),\Moment(t_0+N\cdot h))$ and the approximation $(\Link_N, \Moment_N)$ after $N$ steps of the numerical scheme with step size $h$.

It is readily checked that the SPRK method \eqref{eq:sprk_ode_lie} is consistent in the sense of Theorem~8.1 in \cite[II.8]{hairer_norsett_wanner:solving1} and, hence, has an asymptotic expansion of the form 
\begin{multline}\label{eq:sprk_global}
\|\Link(t_N) - \Link_N\| =\\ e_{\Link,p}(t_N) h^p +  e_{\Link,p+1}(t_N)h^{p+1} + \dotsb + e_{\Link,k}(t_N)h^N+ E_{\Link,h}(\Link,t_N)h^{k+1}.
\end{multline}
Thereby, the $e_{\Link,\nu}(t_N)$ are solutions of some differential equations and $E_{\Link,h}$ is bounded on $[t_0,t_{\text{end}}]$ (accordingly for $\Moment$). Furthermore, it holds $t_N=t_0+N\cdot h$.

As, in addition the SPRK method is symmetric, Theorem~8.10 in \cite[II.8]{hairer_norsett_wanner:solving1} applies, which says, that the expansion \eqref{eq:sprk_global} only contains even powers of $h$:
\begin{align*}
\|\Link(t_N) - \Link_N\| = e_{\Link,2q}(t_N) h^{2q} +  e_{\Link,2q+2}(t_N)h^{2q+2} + \dotsb
\end{align*}

In view of the third order SPRK method with 3 stages defined by the set of coefficients \eqref{eq:sprk_coeffs}, we therefore have a global error
\begin{align*}
\begin{aligned}
\|\Link(t_N) - \Link_N\| &= \mathcal{O}(h^4), & \|\Moment(t_N) - \Moment_N\| &= \mathcal{O}(h^4)
\end{aligned}
\end{align*}
where $(\Link(t),\Moment(t))$ is the exact solution of the equations of motion \eqref{eq:hamilton} and $(\Link_N,\Moment_N)$ are the approximations produced by the SPRK method \eqref{eq:sprk_ode_lie}.

\section{Numerical Results}
We test the SPRK method of order 4 numerically. For this purpose, we simulate an $\SU(2,\C)$ gauge field, used in lattice Quantum Chromodynamics, by means of the Hybrid Monte Carlo method.
The equations of motion \eqref{eq:hamilton} are generated by an Hamiltonian  
\begin{equation*}
 H(\Link,\Moment)=  E_{\mbox{kin}}(\Moment) + S_G(\Link)
\end{equation*}
with kinetic energy $E_{\mbox{kin}}$ and so-called Wilson action $S_G(\Link)$. This is in detail described in \cite[paragraph 7.2.3]{DeGrand:2006}.
Thereby, $\Link = (\link_\nu)_{\nu=1,\dotsc,n}$ is a field of elements of the special unitary Lie group $\SU(2,\C)$, 
and $\Moment = (\moment_\nu)_{\nu=1,\dotsc,n}$ a field of elements of its associated special unitary Lie algebra $\su(2,\C)$. 

In lattice gauge theories, the elements of $\Link$ are called links. Each link $\link_\nu$ has an associated fictitious momentum $p_\nu$ which is a traceless and hermitian 2-by-2 matrix. 
Thus, the momentum $p_\nu$ is connected to traceless and ant-hermitian element $\moment_\nu$ of the Lie algebra $\su(2,\C)$ via a multiplication with the complex $i$: $\Moment = (\moment_\nu = i p_\nu)_{\nu=1,\dotsc,n}$ .

For completeness, note that the function $g_\nu(Y)$ from equation \eqref{eq:hamilton_momentum} is not just evaluated at the lattice point $\nu$ itself but also on some adjacent lattice points called staples.
However, this fact is not important for the derivation of the SPRK method.\\

The simulation is performed on a 2-dimensional lattice of dimension 8x8.
As energy is preserved analytically along the
trajectory, $\langle |\Delta H| \rangle$, 
the mean of the absolute difference between the numerical approximation to the
Hamiltonian at the end of each trajectory of unit length 1
and the initial Hamiltonian, is
an easy to get measure for the numerical approximation error. Figure \ref{fig_num1} reveals a global error $\mathcal{O}(h^4)$ of the SPRK method compared to an error of order $\mathcal{O}(h^2)$ for the Leapfrog scheme. This fact is reflected in Figure \ref{fig_num2},  showing
that the SPRK method is more efficient than Leapfrog with respect to CPU time at smaller energy violations $\langle|\Delta H|\rangle$. For lower error tolerances, however, the numerical effort for fixed-point iteration in the SPRK method becomes visible, and the
efficiency is drastically reduced.
The area preservation mentioned in section \ref{section:eqm} is as expected not met: numerically we have an error of order $4$ for the determinant of the Jacobian $\partial (\Link_N, \Moment_N) / \partial(\Link_0, \Moment_0)$.  

\begin{figure}[h!]
\begin{center}
  \includegraphics[scale=0.6]{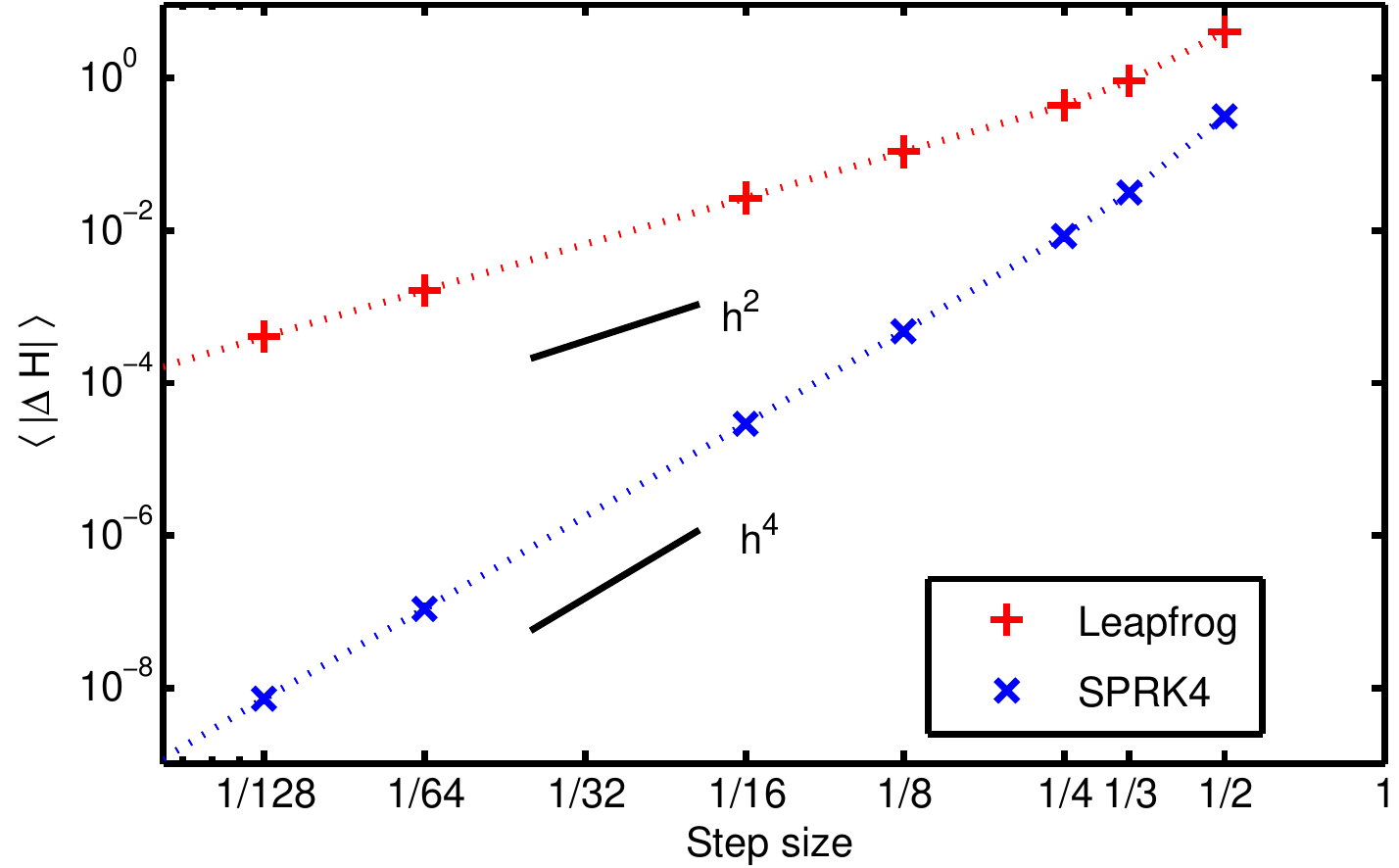}
  \caption{Numerical approximation error of $\langle| \Delta H| \rangle$ for Leapfrog ($+$) and SPRK ($\times$) scheme for different step sizes. The mean of the energy change $|\Delta H|$ along a trajectory with length $1$ is computed from a simulation that is comprised of $5000$ trajectories on a 2-dimensional 8x8 lattice.}\label{fig_num1} 
\end{center}
\end{figure}

\begin{figure}[h!]
\begin{center}
  \includegraphics[scale=0.6]{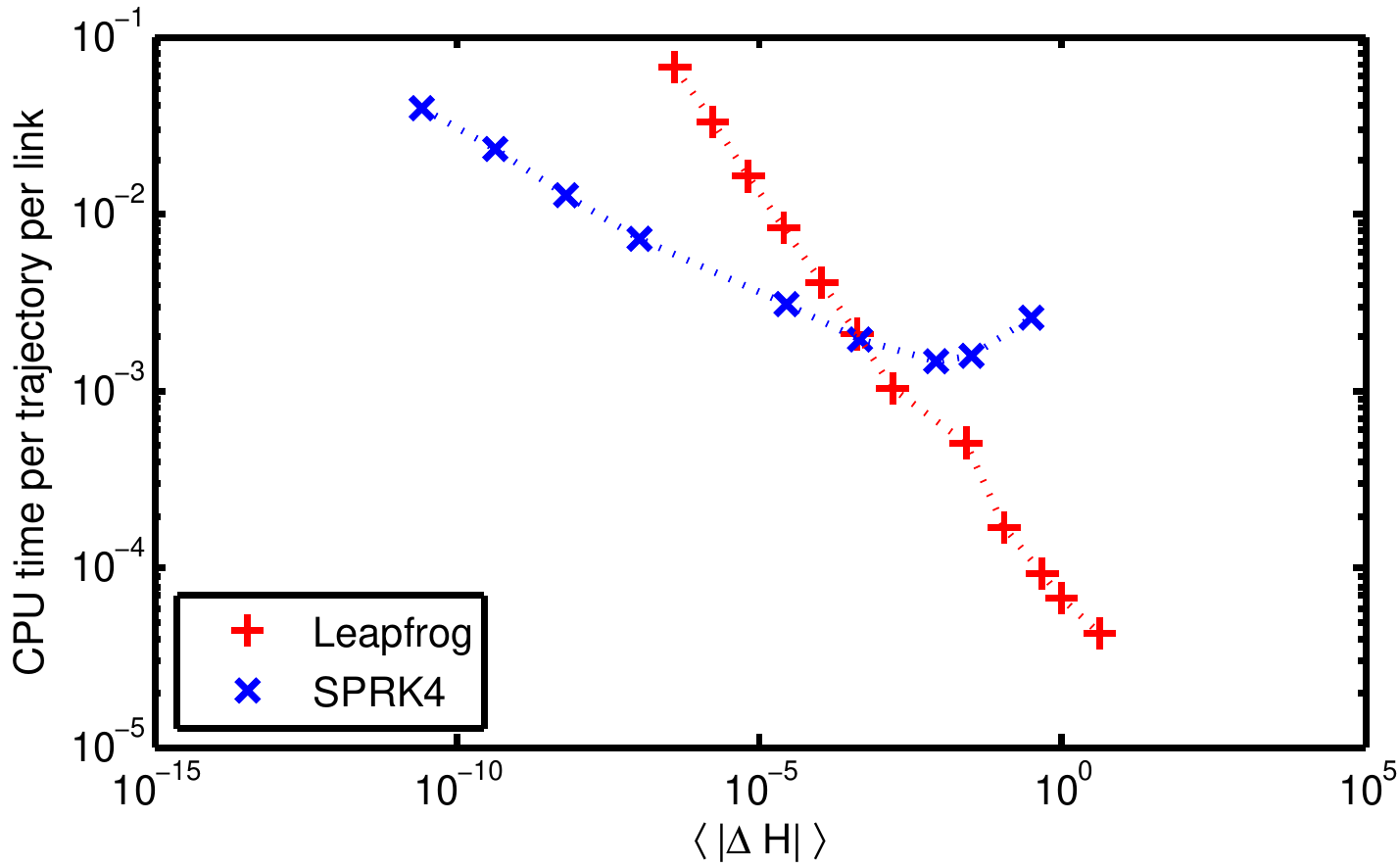}
  \caption{CPU time versus accuracy for Leapfrog ($+$) and SPRK ($\times$). These values are measured in the aforementioned simulation on a 2-dimensional lattice of the size 8x8.}
 \label{fig_num2}
 \end{center}
\end{figure}

\section{Conclusion and Outlook}
In this paper, we have developed symmetric partitioned Runge-Kutta schemes
for matrix differential equations of type~(\ref{eq:hamilton}),
which preserve both Lie-group structure and time-reversibility, and allow
for higher order at the same time. Especially for smaller energy violations (higher error tolerances), 
SPRK schemes have turned out to be more efficient than the Leapfrog scheme.
However, for larger energy violations (lower error tolerances), the fixed-point iteration turned
out to be the computational bottleneck. This drawback has to be attacked
in future works. One idea is to replace the somehow artificial term $\exp(\Omega_1/2)$ introduced to allow for symmetry, which causes a strong
coupling of all components and thus is responsible for a high degree of
non-linearity in system~(\ref{eq:sprk_ode_lie}). As symplecticity is
another desirable property, which simplifies the acceptance step within HMC,
the derivation of symplectic SPRK schemes is a next natural step.

\appendix

\bibliographystyle{elsart-num-sort}

\end{document}